\documentclass[sigconf]{acmart}
\usepackage{pgfplots}
\usepackage{booktabs} 
\usepackage{amsmath}
\usepackage{graphicx}
\usepackage{subcaption}
\usepackage{tikz}
\usepackage{float}
\usepackage[fleqn]{nccmath}
\usepackage{mathtools}
\usepackage[T1]{fontenc}
\usepackage{lmodern}
\setcopyright{rightsretained}

\acmConference[SAMOS'18]{ACM SAMOS conference}{July 2018}{Samos Island, Greece}
\acmYear{2018}
\copyrightyear{2018}

\begin{document}
\title{Single Stream Parallelization of Recurrent Neural Networks
\\ for Low Power and Fast Inference}

\author{Wonyong Sung}

\affiliation{%
  \institution{Seoul National University}
  \city{Seoul}
  \state{Korea}
}
\email{wysung@snu.ac.kr}

\author{Jinhwan Park}
\affiliation{%
  \institution{Seoul National University}
  \city{Seoul}
  \state{Korea}
}
\email{bnoo@snu.ac.kr}

\begin{abstract}
As neural network algorithms show high performance in many
applications, their efficient inference on mobile and embedded systems
are of great interests. When a single stream recurrent neural network (RNN) is executed for a personal user in embedded
systems, it demands a large amount of DRAM accesses because the
network size is usually much bigger than the cache size and the
weights of an RNN are used only once at each time step. We overcome
this problem by parallelizing the algorithm and executing it multiple time steps at a time. This approach also reduces the power consumption by lowering the number of DRAM accesses. QRNN (Quasi Recurrent
Neural Networks) and SRU (Simple Recurrent Unit) based recurrent
neural networks are used for implementation. The experiments for
SRU showed about 300\% and 930\% of speed-up when the numbers
of multi time steps are 4 and 16, respectively, in an ARM CPU based
system.

\end{abstract}

%
%
%


\maketitle

\section{Introduction}

Neural network algorithms show high performance in many machine learning tasks, such as image classification, speech recognition, hand gesture recognition, and machine translation \cite{cho2014properties, graves2005framewise, sak2014long, sundermeyer2012lstm, shin2016dynamic, tang2015document, chan2016listen}. Thus, many future embedded systems, including smartphones and cars, are expected to support many neural network applications. A few different neural network models are used according to the characteristics of the applications. For example, convolutional neural networks are widely used for image recognition \cite{krizhevsky2012imagenet, he2016deep}, while recurrent neural networks are applied to automatic speech recognition (ASR) and machine translation \cite{cho2014properties, graves2005framewise, sak2014long, graves2013speech, cho2014learning}. Many neural network algorithms demand a very large number of arithmetic and memory access operations for real-time operation, and also require very large number of parameters, often exceeding one hundred megabytes (MBs). 

Since the number of parameters is very large compared to the cache size, the overhead of DRAM access is mostly the bottleneck in real-time inference of neural networks on embedded systems. In sever based implementations, batch processing is widely used for lowering the number of DRAM accesses \cite{turchenko2013efficient}. Increasing the batch size lowers the number of DRAM accesses for each stream. However, batch processing can hardly be used in many embedded systems because the application is intended for a single user. One example is an on-device ASR on smartphones. On-device ASR can reduce the delay of response and helps keeping privacy. Although, many researches are being conducted to execute neural networks efficiently with special purpose hardware, there are still DRAM access bottlenecks unless all the parameters are stored on on-chip memory \cite{kim2014x1000}. Considering the available on-chip memory size of most embedded devices, it is still unavoidable to store the weights on DRAM. Thus, it is very needed to constrain the number of DRAM accesses for efficient execution of neural network algorithms. 

Recurrent neural networks are used for sequence processing, and a sequence consists of multiple time steps. Therefore, the number of DRAM accesses can be reduced if multiple time steps are processed at a time. In server based batch processing, multiple streams are processed in parallel to exploit this characteristic. However, for single stream processing, it is very difficult to parallelize the recurrent neural networks because the feedback incurs the dependency problem. Although signal flow graph analysis allows some partial parallelization even for the LSTM (Long Short Term Memory) RNN, which is most widely used, the effect is limited \cite{hwang2015single}. In addition, disconnecting parallelism using data characteristics has also been attempted, but it is difficult to apply this technique to general applications \cite{ouyang2017fast}. Since this problem is due to the nature of recurrent neural network algorithms, it is difficult to overcome this by employing very smart parallelization or special hardware design techniques. 

In this paper, we present a single sequence parallelization of QRNN (Quasi Recurrent Neural Network) and SRU (Simple Recurrent Unit) based recurrent neural networks, and apply multi time step parallelization for efficient implementation on embedded systems. QRNN and SRU algorithms also contain the feed-back structure, but this part can be isolated and takes a small portion of the total computation. This simple feed-back structure allows multi time step parallelization even for a single stream input, and enables efficient implementations reducing DRAM accesses because one weight fetch from DRAM can be used for multiple time steps. QRNN and SRU algorithms are implemented on Intel and ARM CPU based systems, and the execution performances are compared with those of LSTM RNNs.
  
This paper is organized as follows. Section 2 explains the application of RNNs and popular RNN models, including LSTM RNN, QRNN, and SRU. Section 3 presents the strategy for multi time step processing of single stream input. Experimental results are shown in Section 4. Concluding remarks follow in Section 5.

\section{Recurrent neural network algorithms}
A recurrent neural network (RNN) connects the units to form a feedback along a sequence, which allows it to learn dynamic temporal behavior of a time sequence. RNNs can have internal states by either feedback or explicitly using memory cells. This makes them applicable to many sequence recognition tasks. Their applications as well as models that include LSTM, SRU, and QRNN are reviewed in this section. 

\subsection{Application of recurrent neural networks}
\begin{figure}[h!]
  \centering
  \begin{subfigure}[b]{\linewidth}
    \includegraphics[width=\linewidth]{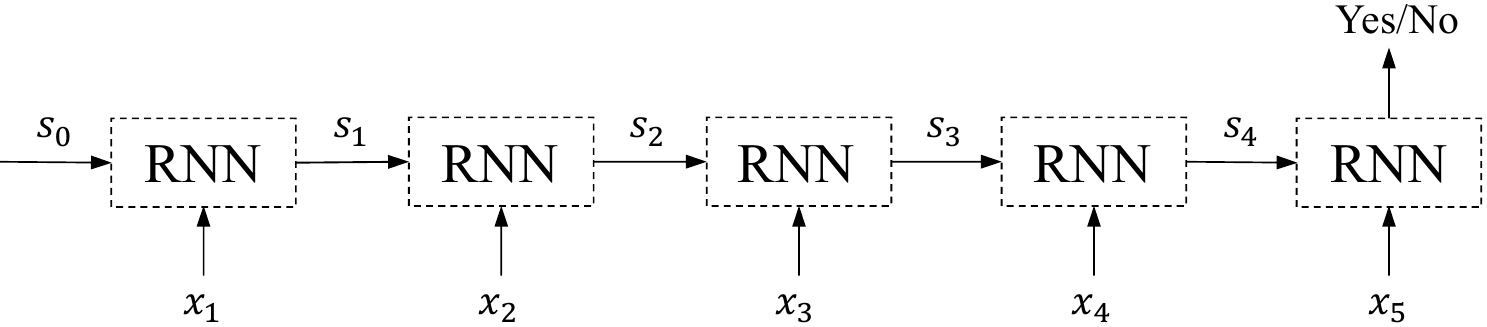}
    \caption{RNN Acceptor.}
  \end{subfigure}
  \begin{subfigure}[b]{\linewidth}
    \includegraphics[width=\linewidth]{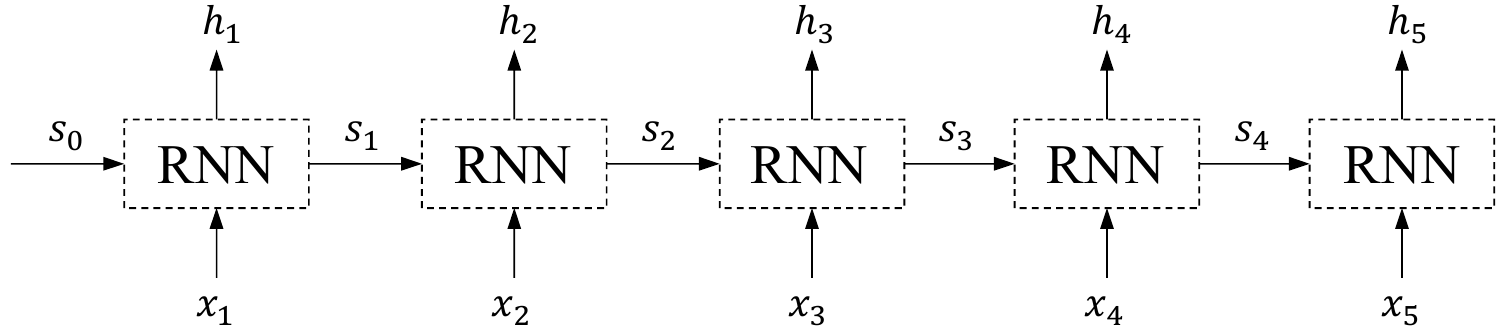}
    \caption{RNN Transducer.}
  \end{subfigure}
    \begin{subfigure}[b]{\linewidth}
    \includegraphics[width=\linewidth]{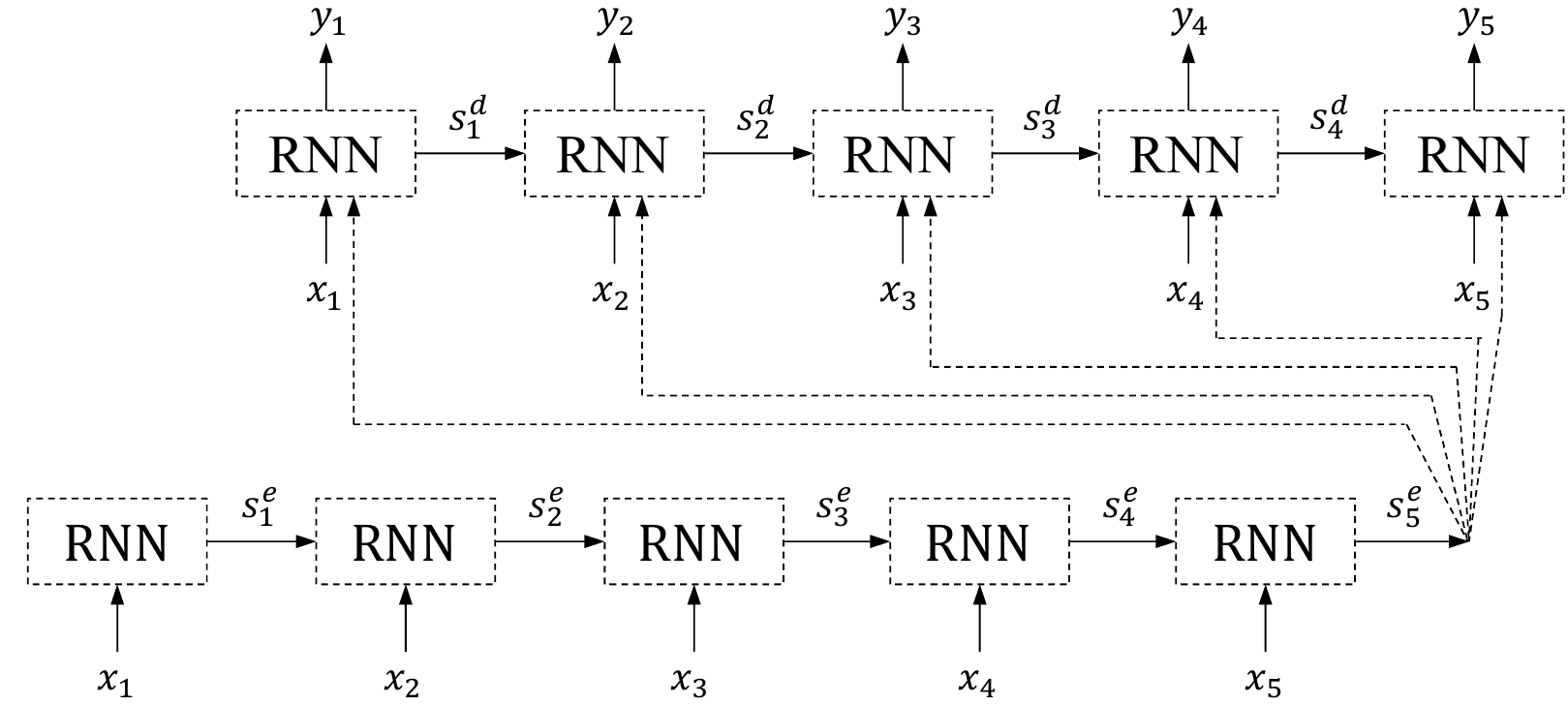}
    \caption{RNN Encoder-Decoder.}
  \end{subfigure}
  \caption{Application of recurrent neural networks.}
  \label{fig:rnn_app}
\end{figure}

RNNs are used for sequence analysis, including text processing, speech recognition, and handwriting recognition. Also, RNN is an indispensable components for sequence generation, such as text generation and foreign language translation.  Three major structures of RNN applications are shown in Fig. \ref{fig:rnn_app}. Fig. \ref{fig:rnn_app} (a) shows the RNN acceptor that receives a whole sequence of input, and then generates an output at the end of the sequence. An application of the RNN acceptor is the sentiment analysis of the input text, such as movie and restaurant reviews \cite{tang2015document}.  Fig. \ref{fig:rnn_app} (b) shows the RNN transducer that receives the input sequence and generates the corresponding output at each time step. The RNN transducer is used for acoustic modeling, language modeling, and so on. Fig. \ref{fig:rnn_app} (c) shows the encoder and decoder model using RNN, where the encoder compresses the input stream and transfers the compressed context to the decoder at the end of the sequence. Thus, the encoder alone is similar to the RNN acceptor. The decoder receives the compressed context as the initial state, and then generates the output as the time step goes. The attention model is an extension of the encoder-decoder architecture \cite{cho2014properties}. The representative application of the encoder-decoder architecture is the language translation. The original text to translate is applied to the encoder, and the translated text is generated from the decoder. 
In many applications, bi-directional RNN models are used, where the sequence flow goes in both directions. The bi-directional RNN can be constructed by combining two RNNs operating at different directions. 

\begin{figure}
\includegraphics[width=\linewidth]{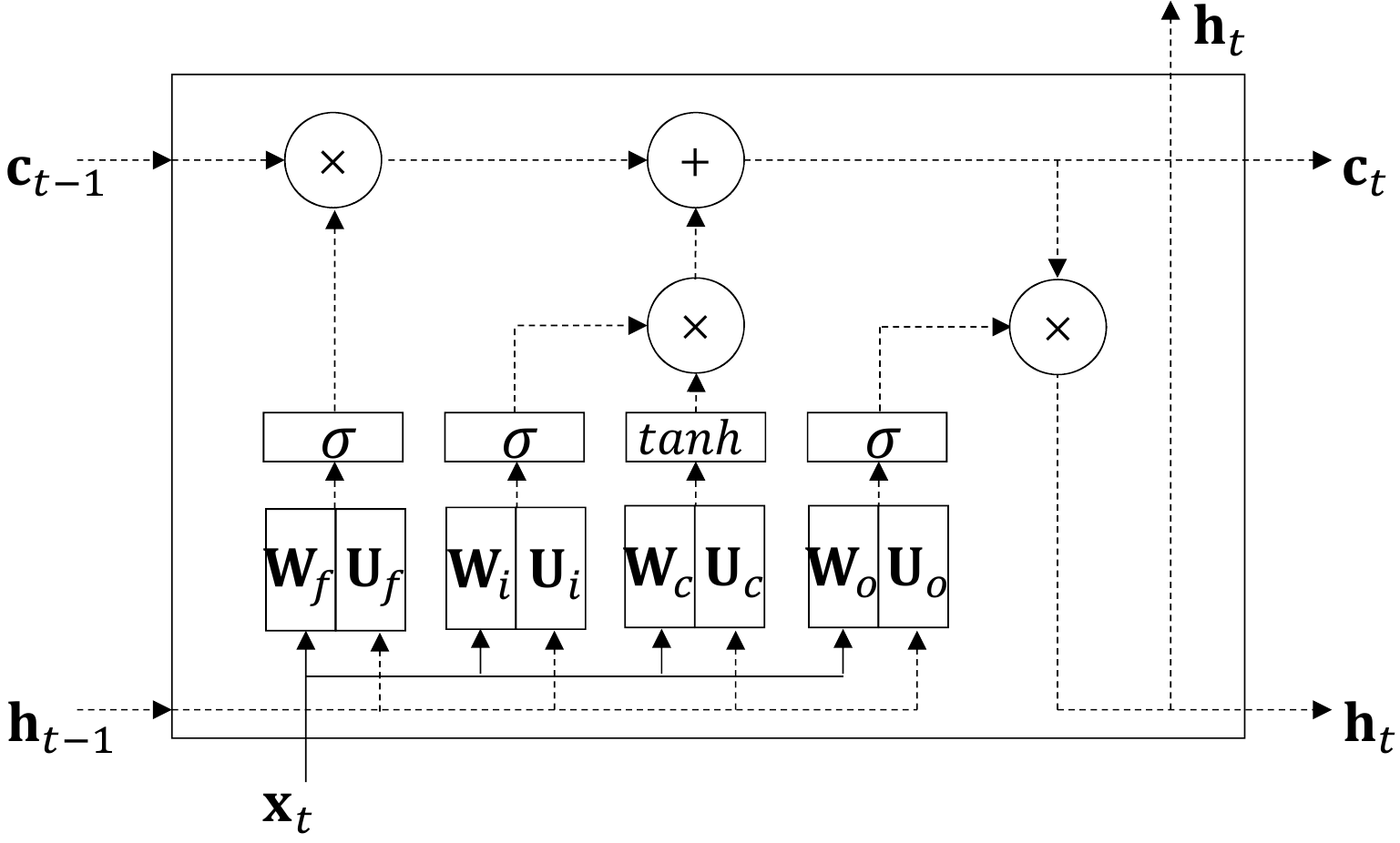}
\caption{Block diagram of LSTM.}
\label{fig:lstm}
\end{figure}

\begin{figure}
\includegraphics[width=\linewidth]{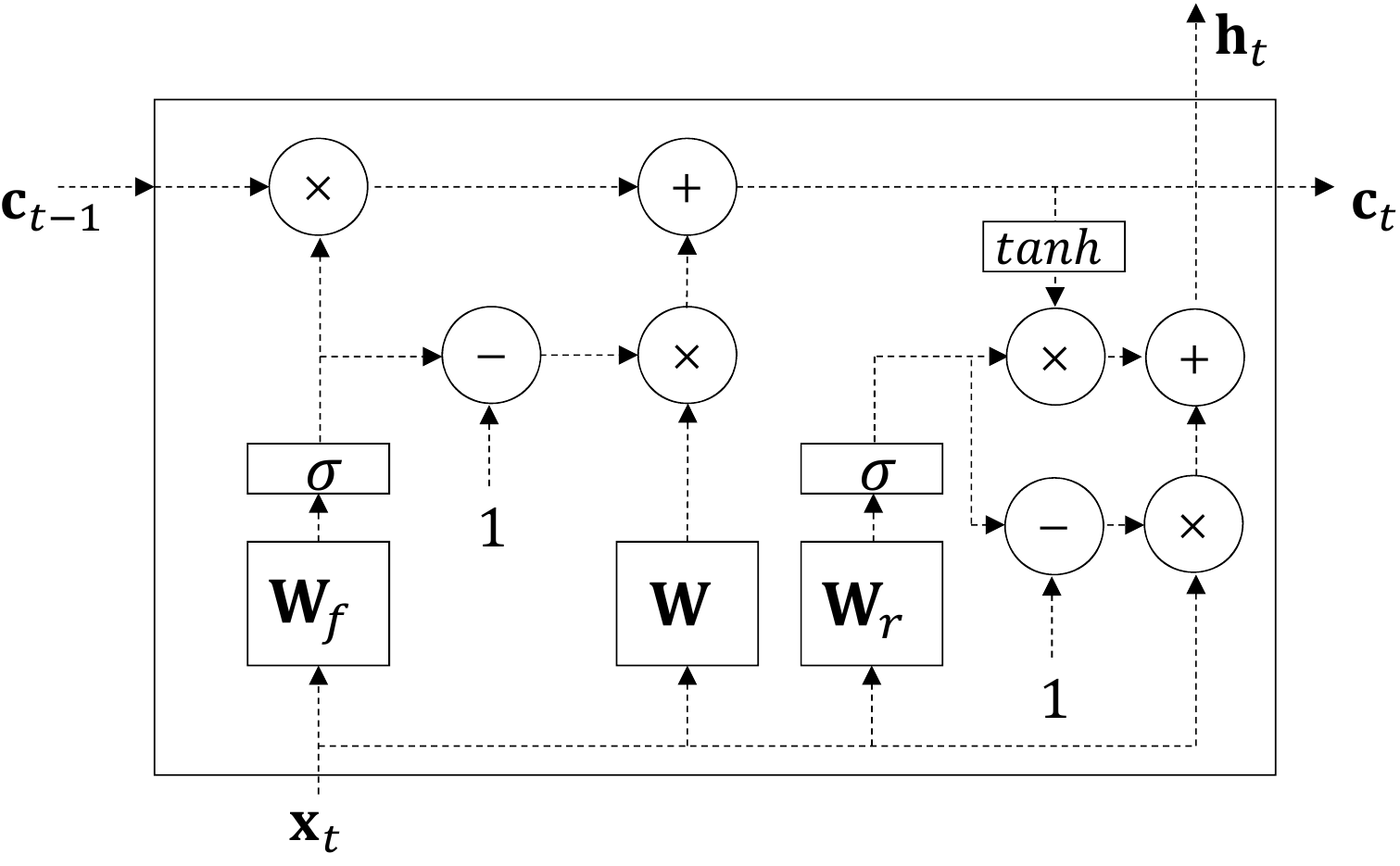}
\caption{Block diagram of SRU.}
\label{fig:sru}
\end{figure}

\subsection{LSTM RNN}
LSTM RNN is most widely used among several RNN models \cite{hochreiter1997long}, and the block diagram is shown in Fig. \ref{fig:lstm}. Also the equations describing LSTM RNN is in Eq. (1). LSTM RNN contains the memory cell, which can store long term information, and whose value is denoted as $\mathbf{c}_t$. The $\mathbf{c}_t$ is propagated to the next time step after being multiplied to the forget gate value, $\mathbf{f}_t$. The output $\mathbf{h}_t$ is generated by using $\mathbf{c}_t$ and the output gate $\mathbf{o}_t$. Note that the $\mathbf{f}_t$ and $\mathbf{o}_t$ are between 0 and 1, and they are formed by transforming both the input $\mathbf{x}_t$ and the previous output $\mathbf{h}_{t-1}$. Thus, without $\mathbf{h}_{t-1}$, it is not possible to compute $\mathbf{f}_{t}$, $\mathbf{i}_{t}$, and $\mathbf{o}_{t}$. As a result, the LSTM equation strictly dictates sequential processing. As indicated in (1), it is possible to precompute $\mathbf{W}_f \mathbf{x}_t$, $\mathbf{W}_i \mathbf{x}_t$, and $\mathbf{W}_o \mathbf{x}_t$ using the multi time processing approach, but the remaining part that depends on $\mathbf{h}_{t-1}$ hinders fully multi time step processing. LSTM RNN approximately demands 8 multiply-vector operations.

\begin{equation}
  \label{equation_lstm}
  \begin{split}
    &\mathbf{f}_t = \sigma\left(\mathbf{W}_f \mathbf{x}_t + \mathbf{U}_f \mathbf{h}_{t-1} + \mathbf{b}_f\right),  \\
    &\mathbf{i}_t = \sigma\left(\mathbf{W}_i \mathbf{x}_t + \mathbf{U}_i \mathbf{h}_{t-1} + \mathbf{b}_i\right), \\
    &\mathbf{o}_t = \sigma\left(\mathbf{W}_o \mathbf{x}_t + \mathbf{U}_o \mathbf{h}_{t-1} + \mathbf{b}_o\right),  \\
    &\mathbf{\hat{c}}_t = \tanh\left(\mathbf{W}_c \mathbf{x}_t + \mathbf{U}_c \mathbf{h}_{t-1} + \mathbf{b}_c\right),  \\
    &\mathbf{c}_t = \mathbf{f}_t \odot \mathbf{c}_{t-1} + \mathbf{i}_t \odot \mathbf{\hat{c}}_t, \\
    &\mathbf{h}_t = \mathbf{o}_t \odot \tanh\left(\mathbf{c}_{t}\right). 
   \end{split}
\end{equation}

\subsection{SRU and QRNN}
The SRU is recently introduced \cite{lei2017training}, and the block diagram is shown in Fig. \ref{fig:sru}. When the SRU is compared with the LSTM, the SRU has only one dependency relation through $\mathbf{c}_{t-1}$, while LSTM depends both $\mathbf{c}_{t-1}$ and $\mathbf{h}_{t-1}$. The SRU can be represented as shown in Eq. (2). Here, we can find that $\mathbf{\hat{x}}_t$, $\mathbf{f}_t$, and $\mathbf{r}_t$ are computed without $\mathbf{h}_{t-1}$.  

\begin{equation}
  \label{equation_sru}
  \begin{split}
    &\mathbf{\hat{x}}_t = \mathbf{W} \mathbf{x}_t ,  \\
    &\mathbf{f}_t = \sigma\left(\mathbf{W}_f \mathbf{x}_t + \mathbf{b}_f\right), \\
    &\mathbf{r}_t = \sigma\left(\mathbf{W}_r \mathbf{x}_t + \mathbf{b}_r\right), \\   
    &\mathbf{c}_t = \mathbf{f}_t \odot \mathbf{c}_{t-1} + \left(1 - \mathbf{f}_t\right) \odot \mathbf{\hat{x}}_t, \\
    &\mathbf{h}_t = \mathbf{r}_t \odot \tanh{\left(\mathbf{c}_{t}\right)} + \left(1 - \mathbf{r}_t\right) \odot \mathbf{x}_t. 
   \end{split}
\end{equation}

The QRNN equation is shown in Eq. (3) \cite{bradbury2016quasi}. When compared to the SRU, the gates are computed using the current and past inputs, but not the past output $\mathbf{h}_{t-1}$. The performances of SRU and QRNN have been studied intensively in recent years. Although the results differ according to the applications, their performances are comparable in many cases when their parameter sizes are similar \cite{lei2017training, bradbury2016quasi}. 

\begin{equation}
  \label{equation_qrnn}
  \begin{split}
    &\mathbf{\hat{x}}_t = \tanh\left(\mathbf{W}^0  \mathbf{x}_t + \mathbf{W}^1  \mathbf{x}_{t-1}\right) ,  \\
    &\mathbf{f}_t = \sigma\left(\mathbf{W}_f^0 \mathbf{x}_t + \mathbf{W}_f^1 \mathbf{x}_{t-1}\right), \\
    &\mathbf{o}_t = \sigma\left(\mathbf{W}_r^0 \mathbf{x}_t + \mathbf{W}_r^1 \mathbf{x}_{t-1}\right), \\   
    &\mathbf{c}_t = \mathbf{f}_t \odot \mathbf{c}_{t-1} + \left(1 - \mathbf{f}_t\right) \odot \mathbf{\hat{x}}_t, \\
    &\mathbf{h}_t = \mathbf{o}_t \odot \tanh{\left(\mathbf{c}_{t}\right)} . 
    \end{split}
\end{equation}

\section{Multi Time Step Parallelization}
RNN computation is executed as a sequence of $\mathbf{h}_0$, $\mathbf{h}_1$, $\mathbf{h}_2$, ... and so on, where $\mathbf{h}_t$ is the output at time step $t$, because evaluation of $\mathbf{h}_t$ needs $\mathbf{h}_{t-1}$ as illustrated in Eq. 1. Multiple time step processing refers concurrent computation of $\mathbf{h}_0$, $\mathbf{h}_1$, ... and $\mathbf{h}_{T-1}$, where $T$ is the block size for parallel processing. In single step processing, the weights, which is usually tens or hundreds of MBs, need to be loaded at each time step, and they are used only once.  Unless the cache size is very large to accommodate all the weights, this incurs a very large number of DRAM accesses. In multi time step processing, a weight is fetched from DRAM and used for processing multiple time steps. Usually, we fetch one row of weight matrix, and use it for computing the output for multiple time steps. As a result, the number of DRAM accesses can be reduced as the number of time steps to process increases. 

\subsection{LSTM}
As illustrated in Eq. (1), the major computation of LSTM network is due to 8 matrix-vector multiplication operations. The matrix values are weights, which are already determined through training, while the vectors are either the input $\mathbf{x}_t$ or the previous output $\mathbf{h}_{t-1}$. As for the matrix-vector multiplication with the input $\mathbf{x}_t$, there is no difficulty in multiple time step processing. But, the matrix-vector multiplication with the previous output $\mathbf{h}_{t-1}$ cannot be conducted employing the multiple time step fashion because of the dependency problem. Thus, even if we precompute the matrix-vector computation with the input vector $\mathbf{x}_t$, the number of DRAM accesses can be reduced just up to a half, when compared with the single time step processing. 

\subsection{SRU and QRNN}
SRU computes the output through three matrix-vector multiplication operations as shown in Eq. (2). Here we can find that the computation of weighted input $\mathbf{\hat{x}}_t$, forget gate $\mathbf{f}_t$, and output gate $\mathbf{r}_t$ can be conducted using only the weight matrices and the input $\mathbf{x}_t$. There is no dependency relation with the output $\mathbf{h}_{t-1}$. 
In SRU, there is a dependency loop propagating the memory cell value $\mathbf{c}_t$, which is, however, vector element wise operations and demands much less operations compared to matrix-vector multiplications. Also, the vector element wise operation can be conducted up to the parallelism of the layer width of RNN, which usually have the range of 128 to 1024. Thus, the computation of $\mathbf{c}_t$ can be conducted using SIMD or multi-thread operations. The computation of output $\mathbf{h}_t$ has no dependency constraints when its input $\mathbf{r}_t$ and $\mathbf{c}_t$ are all given. 

The QRNN equation is shown in Eq. (3). When compared to the SRU, the gates are computed using the current and past inputs, but not the past output $\mathbf{h}_{t-1}$. Thus, this structure can also be multi time step parallelized at the same way.
When employing the multi time step approach, the forget gate signal,  $\mathbf{f}_t$, can be computed as shown in Eq. (4). [$\mathbf{f}_0$, $\mathbf{f}_1$, ..., $\mathbf{f}_T$] are computed at a time as Eq. (4) shows. Here, we can use the matrix-matrix multiplication. Note that $\mathbf{f}_t$ is a column vector whose size is usually between 128 and 2,048 in many RNN applications. Matrix-matrix multiplication uses the weights several times, thus it demands far less external memory accesses for each arithmetic \cite{quinn2003parallel}.

\begin{equation}
  \label{equation_parallel}
  \begin{split}
\\
\begin{bmatrix}
     &  \\
    \mathbf{f}_0 &\mathbf{f}_1 & ... &  \mathbf{f}_T  \\
     & 
\end{bmatrix}
=
\begin{bmatrix}
    & &\\
 & \mathbf{W}_f& \\
    & &
\end{bmatrix}
\begin{bmatrix}
     &  \\
    \mathbf{x}_0 &\mathbf{x}_1 & ... &  \mathbf{x}_T  \\
     & 
\end{bmatrix}\\
\\
    \end{split}
\end{equation}

\begin{figure}[h]
  \centering
  \begin{subfigure}[b]{\linewidth}
    \includegraphics[width=\linewidth]{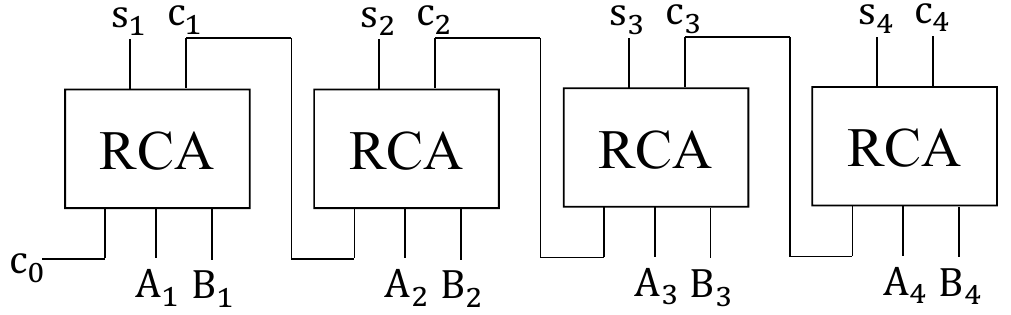}
    \caption{Ripple carry adder.}
  \end{subfigure}
  \begin{subfigure}[b]{\linewidth}
    \includegraphics[width=\linewidth]{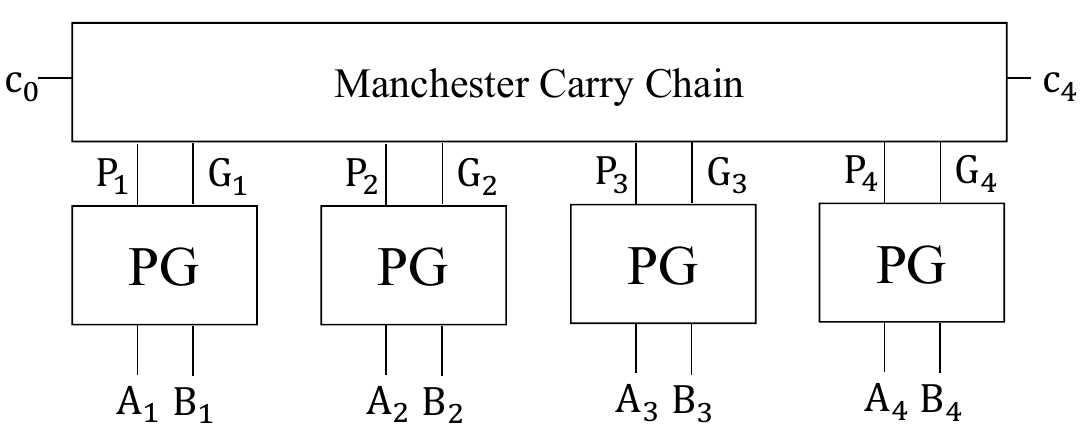}
    \caption{Manchester carry chain adder.}
  \end{subfigure}
  \caption{Ripple carry adder and Manchester carry chain adder.}
  \label{fig:adder}
\end{figure}

The proposed idea is similar to the Manchester carry chain adder shown in Fig. \ref{fig:adder}. In the Manchester carry chain, the propagation and generation signals are pre-computed only using the input. Then, the carry is propagated very fast using the propagation and generation signals \cite{weste1994principles}. 

\section{Experimental results}
We have measured the execution time of LSTM RNN, SRU, and QRNN models on Intel Core i7-3930K 3.2GHz CPU and Nvidia Denver2 ARMv8 64-bit 2.0GHz CPU. Small and large RNN models are used for the experiments. As for the small model, the LSTM RNN has the cell width of 350, while the SRU has the width of 512, and the number of parameters are approximately 1M.  The large model LSTM RNN has the cell width of 700, and that of SRU RNN has the width of 1024, by which the parameter size of both models are comparable, approximately 3M.

 The Intel CPU has 32KB L1 data cache and instruction cache, 256KB L2 cache, and 12,288KB L3 cache. The ARM CPU has 32KB L1 data cache, 48KB instruction cache, and 2,048KB L2 cache. The program was written in C++ language, and used the BLAS for optimum execution of matrix-vector and matrix-matrix multiplications. Intel Math Kernel Library (MKL) \cite{intel2015mklblas} and OpenBLAS \cite{xianyi2012openblas} are used for Intel CPU and ARM CPU, respectively. The highest compiler optimization level, O2, is used for this experiments. The time is measured while processing 1,024 input samples. 

\begin{table}
  \caption{Execution time (msec) of small model RNNs on the Intel CPU for processing 1,024 input samples.}
  \label{tab:small_Intel}
  \begin{tabular}{lll}
    \toprule
    Model&Execution Time&Speed-up\\
    \midrule
    LSTM & 673.667& -\\
    SRU-1 & 475.43& 100\% \\
    SRU-2 & 288.729& 164.7\%\\    
    SRU-4 & 197.765& 240.4\%\\
    SRU-8 & 153.39& 309.9\%\\
    SRU-16 & 129.591& 366.9\%\\
    SRU-32 & 118.247& 402.1\%\\            
    SRU-64 & 96.302& 493.7\%\\            
    SRU-128 & 93.219& 510.0\%\\                        
  \bottomrule
\end{tabular}
\end{table}

\begin{table}
  \caption{Execution time (msec) of large model RNNs on the Intel CPU.}
  \label{tab:large_Intel}
  \begin{tabular}{lll}
    \toprule
    Model&Execution Time&Speed-up\\
    \midrule
    LSTM & 2359.94& -\\
    SRU-1 & 1880.63& 100\% \\
    SRU-2 & 1104.22& 170.3\%\\    
    SRU-4 & 715.919& 262.6\%\\
    SRU-8 & 523.264& 359.4\%\\
    SRU-16 & 437.565& 429.7\%\\
    SRU-32 & 375.647& 500.6\%\\            
    SRU-64 & 335.64& 560.3\%\\            
    SRU-128 & 320.121& 587.4\%\\                        
  \bottomrule
\end{tabular}
\end{table}

\begin{table}
  \caption{Execution time (msec) of small model RNNs on the ARM CPU.}
  \label{tab:small_ARM}
  \begin{tabular}{lll}
    \toprule
    Model&Execution Time&Speed-up\\
    \midrule
    LSTM & 1522.3& -\\
    SRU-1 & 902.736& 100\% \\
    SRU-2 & 484.474& 186.3\%\\    
    SRU-4 & 274.82& 328.5\%\\
    SRU-8 & 172.856& 522.2\%\\
    SRU-16 & 108.414& 832.6\%\\
    SRU-32 & 85.6596& 1053.8\%\\            
    SRU-64 & 96.1196& 939.1\%\\            
    SRU-128 & 93.3887& 966.6\%\\                        
  \bottomrule
\end{tabular}
\end{table} 

\begin{table}
  \caption{Execution time (msec) of large model RNNs on the ARM CPU.}
  \label{tab:large_ARM}
  \begin{tabular}{lll}
    \toprule
    Model&Execution Time&Speed-up\\
    \midrule
    LSTM & 4583.75& -\\
    SRU-1 & 3652.59& 100\% \\
    SRU-2 & 1925.07& 189.7\%\\    
    SRU-4 & 1078.03& 338.8\%\\
    SRU-8 & 634.951& 575.3\%\\
    SRU-16 & 392.163& 931.4\%\\
    SRU-32 & 288.659& 1265.4\%\\            
    SRU-64 & 275.078& 1327.8\%\\            
    SRU-128 & 275.658& 1325.0\%\\                        
  \bottomrule
\end{tabular}
\end{table} 

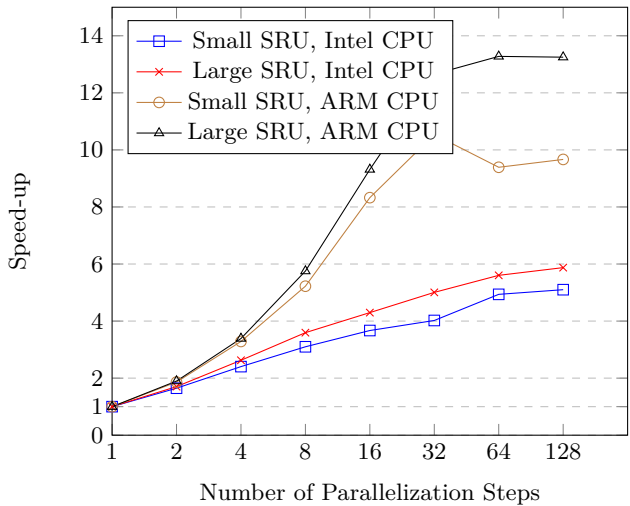
\begin{figure}[h]
\begin{tikzpicture}
\begin{axis}[
    title={},
    xmode=log,
    log basis x=2,
    xlabel={Number of Parallelization Steps},
    ylabel={Speed-up},
    xmin=1, xmax=256,
    ymin=0, ymax=15,
    xtick={1,2,4,8,16,32,64,128},
    xticklabels={1,2,4,8,16,32,64,128},
    ytick={0,1,2,4,6,8,10,12,14},
    legend pos=north west,
    ymajorgrids=true,
    grid style=dashed,
]

\addplot[
    color=blue,
    mark=square,
    ]
    coordinates {
    (1,1)(2,1.647)(4,2.404)(8,3.099)(16,3.669)(32,4.021)(64,4.937)(128,5.100)
    };
\addplot[
    color=red,
    mark=x,
    ]
    coordinates {
    (1,1)(2,1.703)(4,2.626)(8,3.594)(16,4.297)(32,5.006)(64,5.603)(128,5.874)
    };
\addplot[
    color=brown,
    mark=o,
    ]
    coordinates {
    (1,1)(2,1.863)(4,3.285)(8,5.222)(16,8.326)(32,10.538)(64,9.391)(128,9.666)
    };
\addplot[
    color=black,
    mark=triangle,
    ]
    coordinates {
    (1,1)(2,1.897)(4,3.388)(8,5.753)(16,9.314)(32,12.654)(64,13.278)(128,13.250)
    };
    \legend{{Small SRU, Intel CPU}, {Large SRU, Intel CPU}, {Small SRU, ARM CPU}, {Large SRU, ARM CPU}}
\end{axis}
\end{tikzpicture}
\caption{Relative speed-up of SRU according to the number of parallelization steps.} \label{fig:sru_plot}
\end{figure}

Table \ref{tab:small_Intel} gives the execution time and the speed-up of the small model RNNs on the Intel CPU.  The LSTM with single time step and the SRU with multiple time step parallelization are shown. `SRU-$n$' refers to the SRU execution with $n$-step parallelization. The speed-up of SRU-$n$ is measured on the basis of the SRU-1. Here, we can find almost 400\% of speed-up when 32 step parallelization is employed. A similar result can be found for the large model RNNs. Table \ref{tab:large_Intel} shows the execution time and the speed-up of large model RNNs on the CPU. The speed-up is about 500\% when 32 step parallelization is used. 
   
The experimental results on ARM CPU are given on Table 3 and 4 for the small and large models, respectively. When executing the small model on the ARM CPU, the speed-up is nearly 1,000\% when the number of parallelization steps is 32. The speed-up of large model RNN is very impressive, more than 1,250\%.

\begin{table}
  \caption{Execution time (msec) of small QRNNs on the Intel CPU.}
  \label{tab:small_qrnn_Intel}
  \begin{tabular}{lll}
    \toprule
    Model&Execution Time&Speed-up\\
    \midrule
    QRNN-1 & 1034.77& 100\% \\
    QRNN-2 & 558.107& 185.4\%\\    
    QRNN-4 & 376.691& 274.7\%\\
    QRNN-8 & 285.414& 362.5\%\\
    QRNN-16 & 239.941& 431.2\%\\
    QRNN-32 & 216.77& 477.3\%\\            
    QRNN-64 & 173.527& 596.3\%\\            
    QRNN-128 & 167.381& 618.2\%\\                        
  \bottomrule
\end{tabular}
\end{table}

\begin{table}
  \caption{Execution time (msec) of large QRNNs on the Intel CPU.}
  \label{tab:large_qrnn_Intel}
  \begin{tabular}{lll}
    \toprule
    Model&Execution Time&Speed-up\\
    \midrule
    QRNN-1 & 3862.67& 100\% \\
    QRNN-2 & 2194.5& 176.0\%\\    
    QRNN-4 & 1413.61& 273.2\%\\
    QRNN-8 & 1020.05& 378.7\%\\
    QRNN-16 & 834.649& 462.8\%\\
    QRNN-32 & 711.423& 542.9\%\\            
    QRNN-64 & 631.667& 611.5\%\\            
    QRNN-128 & 600.772& 643.0\%\\                        
  \bottomrule
\end{tabular}
\end{table}

\begin{table}
  \caption{Execution time (msec) of small QRNNs on the ARM CPU.}
  \label{tab:smal_qrnn_arm}
  \begin{tabular}{lll}
    \toprule
    Model&Execution Time&Speed-up\\
    \midrule
    QRNN-1 & 1580.58& 100\% \\
    QRNN-2 & 830.659& 190.3\%\\    
    QRNN-4 & 461.075& 342.8\%\\
    QRNN-8 & 323.815& 488.1\%\\
    QRNN-16 & 197.612& 799.8\%\\
    QRNN-32 & 143.158& 1104.9\%\\            
    QRNN-64 & 140.108& 1128.1\%\\            
    QRNN-128 & 142.536& 1108.9\%\\                        
  \bottomrule
\end{tabular}
\end{table}

\begin{table}
  \caption{Execution time (msec) of large QRNNs on the ARM CPU.}
  \label{tab:smal_qrnn_arm}
  \begin{tabular}{lll}
    \toprule
    Model&Execution Time&Speed-up\\
    \midrule
    QRNN-1 & 6467.72& 100\% \\
    QRNN-2 & 3356.7& 192.6\%\\    
    QRNN-4 & 1844.29& 350.6\%\\
    QRNN-8 & 1253.13& 516.1\%\\
    QRNN-16 & 712.439& 907.8\%\\
    QRNN-32 & 475.433& 1360.3\%\\            
    QRNN-64 & 469.515& 1377.5\%\\            
    QRNN-128 & 450.848& 1434.6\%\\                        
  \bottomrule
\end{tabular}
\end{table}

\begin{figure}
\begin{tikzpicture}
\begin{axis}[
    title={},
    xmode=log,
    log basis x=2,
    xlabel={Number of Parallelization Steps},
    ylabel={Speed-up},
    xmin=1, xmax=256,
    ymin=0, ymax=15,
    xtick={1,2,4,8,16,32,64,128},
    xticklabels={1,2,4,8,16,32,64,128},
    ytick={0,1,2,4,6,8,10,12,14},
    legend pos=north west,
    ymajorgrids=true,
    grid style=dashed,
]

\addplot[
    color=blue,
    mark=square,
    ]
    coordinates {
    (1,1)(2,1.854)(4,2.747)(8,3.625)(16,4.312)(32,4.773)(64,5.963)(128,6.182)
    };
\addplot[
    color=red,
    mark=x,
    ]
    coordinates {
    (1,1)(2,1.760)(4,2.732)(8,3.787)(16,4.628)(32,5.429)(64,6.115)(128,6.430)
    };
\addplot[
    color=brown,
    mark=o,
    ]
    coordinates {
    (1,1)(2,1.903)(4,3.428)(8,4.881)(16,7.998)(32,11.049)(64,11.281)(128,11.089)
    };
\addplot[
    color=black,
    mark=triangle,
    ]
    coordinates {
    (1,1)(2,1.926)(4,3.506)(8,5.161)(16,9.078)(32,13.603)(64,13.775)(128,14.346)
    };
    \legend{{Small QRNN, Intel CPU}, {Large QRNN, Intel CPU}, {Small QRNN, ARM CPU}, {Large QRNN, ARM CPU}}
\end{axis}
\end{tikzpicture}
\caption{Relative speed-up of QRNN according to the number of parallelization steps.} \label{fig:qrnn_plot}
\end{figure}
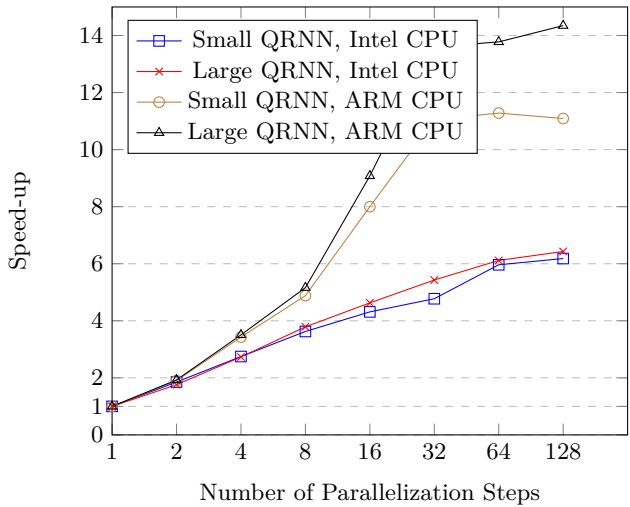

 Fig. \ref{fig:sru_plot} summarizes the speed-up of SRU according to the number of parallelization steps. We can find that the benefit of multiple time step parallelization is bigger in ARM based systems when compared to Intel CPU based computer.  This is because the benefit of reduced DRAM access due to multi time step parallelization becomes more prominent when the computer system has a poor memory system, such as low bandwidth DRAM and small cache size. Also, the larger RNN model that needs more parameters shows higher speed-up compared to the small one.

The experimental results for QRNN are shown in Table 5-8. Here, we can find the similar trends.  We can find higher speed-up values in the ARM based system. Fig. \ref{fig:qrnn_plot} shows the speed-up of QRNN.

\section{Concluding Remarks}
RNN execution demands a large memory bandwidth, especially when only a single stream input data is processed. We have reduced the number of DRAM accesses by parallelizing the RNNs in the time domain. In this approach, one weight fetch is used for execution in multiple time steps, thus the efficiency increases as the number of time steps to parallelize grows. This parallelization technique is applied to simple recurrent units (SRUs) and quasi-RNNs (QRNNs), which are both recently developed models with a simple recurrence structure. The experiments were conducted on Intel CPU and ARM CPU based systems.  We achieved a speed-up of more than 500\% at the Intel CPU based system, and that of more than 1,250\% at the ARM CPU based system. This technique can be utilized for high speed inference of RNNs on VLSI or GPUs (Graphics Processing Units).

\begin{acks}
 This work is supported in part by the Brain Korea
21 Plus Project and the National Research Foundation of Korea
(NRF) grant funded by the Korea government (MSIP)
(No. 2015R1A2A1A10056051). This work is also supported by Samsung
Advanced Institute of Technology.

\end{acks}

\bibliographystyle{ACM-Reference-Format}

\bibliography{sample-bibliography}


\begin{thebibliography}{22}


\ifx \showCODEN    \undefined \def \showCODEN     #1{\unskip}     \fi
\ifx \showDOI      \undefined \def \showDOI       #1{#1}\fi
\ifx \showISBNx    \undefined \def \showISBNx     #1{\unskip}     \fi
\ifx \showISBNxiii \undefined \def \showISBNxiii  #1{\unskip}     \fi
\ifx \showISSN     \undefined \def \showISSN      #1{\unskip}     \fi
\ifx \showLCCN     \undefined \def \showLCCN      #1{\unskip}     \fi
\ifx \shownote     \undefined \def \shownote      #1{#1}          \fi
\ifx \showarticletitle \undefined \def \showarticletitle #1{#1}   \fi
\ifx \showURL      \undefined \def \showURL       {\relax}        \fi
\providecommand\bibfield[2]{#2}
\providecommand\bibinfo[2]{#2}
\providecommand\natexlab[1]{#1}
\providecommand\showeprint[2][]{arXiv:#2}

\bibitem[\protect\citeauthoryear{Bradbury, Merity, Xiong, and Socher}{Bradbury
  et~al\mbox{.}}{2016}]%
        {bradbury2016quasi}
\bibfield{author}{\bibinfo{person}{James Bradbury}, \bibinfo{person}{Stephen
  Merity}, \bibinfo{person}{Caiming Xiong}, {and} \bibinfo{person}{Richard
  Socher}.} \bibinfo{year}{2016}\natexlab{}.
\newblock \showarticletitle{Quasi-recurrent neural networks}.
\newblock \bibinfo{journal}{\emph{arXiv preprint arXiv:1611.01576}}
  (\bibinfo{year}{2016}).
\newblock


\bibitem[\protect\citeauthoryear{Chan, Jaitly, Le, and Vinyals}{Chan
  et~al\mbox{.}}{2016}]%
        {chan2016listen}
\bibfield{author}{\bibinfo{person}{William Chan}, \bibinfo{person}{Navdeep
  Jaitly}, \bibinfo{person}{Quoc Le}, {and} \bibinfo{person}{Oriol Vinyals}.}
  \bibinfo{year}{2016}\natexlab{}.
\newblock \showarticletitle{Listen, attend and spell: A neural network for
  large vocabulary conversational speech recognition}. In
  \bibinfo{booktitle}{\emph{Acoustics, Speech and Signal Processing (ICASSP),
  2016 IEEE International Conference on}}. IEEE, \bibinfo{pages}{4960--4964}.
\newblock


\bibitem[\protect\citeauthoryear{Cho, van Merri{\"e}nboer, Bahdanau, and
  Bengio}{Cho et~al\mbox{.}}{2014a}]%
        {cho2014properties}
\bibfield{author}{\bibinfo{person}{Kyunghyun Cho}, \bibinfo{person}{Bart van
  Merri{\"e}nboer}, \bibinfo{person}{Dzmitry Bahdanau}, {and}
  \bibinfo{person}{Yoshua Bengio}.} \bibinfo{year}{2014}\natexlab{a}.
\newblock \showarticletitle{On the Properties of Neural Machine Translation:
  Encoder--Decoder Approaches}.
\newblock \bibinfo{journal}{\emph{Syntax, Semantics and Structure in
  Statistical Translation}} (\bibinfo{year}{2014}), \bibinfo{pages}{103}.
\newblock


\bibitem[\protect\citeauthoryear{Cho, van Merrienboer, Gulcehre, Bahdanau,
  Bougares, Schwenk, and Bengio}{Cho et~al\mbox{.}}{2014b}]%
        {cho2014learning}
\bibfield{author}{\bibinfo{person}{Kyunghyun Cho}, \bibinfo{person}{Bart van
  Merrienboer}, \bibinfo{person}{Caglar Gulcehre}, \bibinfo{person}{Dzmitry
  Bahdanau}, \bibinfo{person}{Fethi Bougares}, \bibinfo{person}{Holger
  Schwenk}, {and} \bibinfo{person}{Yoshua Bengio}.}
  \bibinfo{year}{2014}\natexlab{b}.
\newblock \showarticletitle{Learning Phrase Representations using RNN
  Encoder--Decoder for Statistical Machine Translation}. In
  \bibinfo{booktitle}{\emph{Proceedings of the 2014 Conference on Empirical
  Methods in Natural Language Processing (EMNLP)}}.
  \bibinfo{pages}{1724--1734}.
\newblock


\bibitem[\protect\citeauthoryear{Graves, Mohamed, and Hinton}{Graves
  et~al\mbox{.}}{2013}]%
        {graves2013speech}
\bibfield{author}{\bibinfo{person}{Alex Graves}, \bibinfo{person}{Abdel-rahman
  Mohamed}, {and} \bibinfo{person}{Geoffrey Hinton}.}
  \bibinfo{year}{2013}\natexlab{}.
\newblock \showarticletitle{Speech recognition with deep recurrent neural
  networks}. In \bibinfo{booktitle}{\emph{Acoustics, speech and signal
  processing (icassp), 2013 ieee international conference on}}. IEEE,
  \bibinfo{pages}{6645--6649}.
\newblock


\bibitem[\protect\citeauthoryear{Graves and Schmidhuber}{Graves and
  Schmidhuber}{2005}]%
        {graves2005framewise}
\bibfield{author}{\bibinfo{person}{Alex Graves} {and}
  \bibinfo{person}{J{\"u}rgen Schmidhuber}.} \bibinfo{year}{2005}\natexlab{}.
\newblock \showarticletitle{Framewise phoneme classification with bidirectional
  LSTM and other neural network architectures}.
\newblock \bibinfo{journal}{\emph{Neural Networks}} \bibinfo{volume}{18},
  \bibinfo{number}{5-6} (\bibinfo{year}{2005}), \bibinfo{pages}{602--610}.
\newblock


\bibitem[\protect\citeauthoryear{He, Zhang, Ren, and Sun}{He
  et~al\mbox{.}}{2016}]%
        {he2016deep}
\bibfield{author}{\bibinfo{person}{Kaiming He}, \bibinfo{person}{Xiangyu
  Zhang}, \bibinfo{person}{Shaoqing Ren}, {and} \bibinfo{person}{Jian Sun}.}
  \bibinfo{year}{2016}\natexlab{}.
\newblock \showarticletitle{Deep residual learning for image recognition}. In
  \bibinfo{booktitle}{\emph{Proceedings of the IEEE conference on computer
  vision and pattern recognition}}. \bibinfo{pages}{770--778}.
\newblock


\bibitem[\protect\citeauthoryear{Hochreiter and Schmidhuber}{Hochreiter and
  Schmidhuber}{1997}]%
        {hochreiter1997long}
\bibfield{author}{\bibinfo{person}{Sepp Hochreiter} {and}
  \bibinfo{person}{J{\"u}rgen Schmidhuber}.} \bibinfo{year}{1997}\natexlab{}.
\newblock \showarticletitle{Long short-term memory}.
\newblock \bibinfo{journal}{\emph{Neural computation}} \bibinfo{volume}{9},
  \bibinfo{number}{8} (\bibinfo{year}{1997}), \bibinfo{pages}{1735--1780}.
\newblock


\bibitem[\protect\citeauthoryear{Hwang and Sung}{Hwang and Sung}{2015}]%
        {hwang2015single}
\bibfield{author}{\bibinfo{person}{Kyuyeon Hwang} {and}
  \bibinfo{person}{Wonyong Sung}.} \bibinfo{year}{2015}\natexlab{}.
\newblock \showarticletitle{Single stream parallelization of generalized
  LSTM-like RNNs on a GPU}. In \bibinfo{booktitle}{\emph{Acoustics, Speech and
  Signal Processing (ICASSP), 2015 IEEE International Conference on}}. IEEE,
  \bibinfo{pages}{1047--1051}.
\newblock


\bibitem[\protect\citeauthoryear{Intel}{Intel}{2015}]%
        {intel2015mklblas}
\bibfield{author}{\bibinfo{person}{Intel}.} \bibinfo{year}{2015}\natexlab{}.
\newblock \showarticletitle{Intel Math Kernel Library Developer Reference}.
\newblock \bibinfo{journal}{\emph{URL:
  https://software.intel.com/en-us/articles/mkl-reference-manual}}
  (\bibinfo{year}{2015}).
\newblock


\bibitem[\protect\citeauthoryear{Kim, Hwang, and Sung}{Kim
  et~al\mbox{.}}{2014}]%
        {kim2014x1000}
\bibfield{author}{\bibinfo{person}{Jonghong Kim}, \bibinfo{person}{Kyuyeon
  Hwang}, {and} \bibinfo{person}{Wonyong Sung}.}
  \bibinfo{year}{2014}\natexlab{}.
\newblock \showarticletitle{X1000 real-time phoneme recognition VLSI using
  feed-forward deep neural networks}. In \bibinfo{booktitle}{\emph{Acoustics,
  Speech and Signal Processing (ICASSP), 2014 IEEE International Conference
  on}}. IEEE, \bibinfo{pages}{7510--7514}.
\newblock


\bibitem[\protect\citeauthoryear{Krizhevsky, Sutskever, and Hinton}{Krizhevsky
  et~al\mbox{.}}{2012}]%
        {krizhevsky2012imagenet}
\bibfield{author}{\bibinfo{person}{Alex Krizhevsky}, \bibinfo{person}{Ilya
  Sutskever}, {and} \bibinfo{person}{Geoffrey~E Hinton}.}
  \bibinfo{year}{2012}\natexlab{}.
\newblock \showarticletitle{Imagenet classification with deep convolutional
  neural networks}. In \bibinfo{booktitle}{\emph{Advances in neural information
  processing systems}}. \bibinfo{pages}{1097--1105}.
\newblock


\bibitem[\protect\citeauthoryear{Lei and Zhang}{Lei and Zhang}{2017}]%
        {lei2017training}
\bibfield{author}{\bibinfo{person}{Tao Lei} {and} \bibinfo{person}{Yu Zhang}.}
  \bibinfo{year}{2017}\natexlab{}.
\newblock \showarticletitle{Training RNNs as Fast as CNNs}.
\newblock \bibinfo{journal}{\emph{arXiv preprint arXiv:1709.02755}}
  (\bibinfo{year}{2017}).
\newblock


\bibitem[\protect\citeauthoryear{Ouyang, Yin, and Wei}{Ouyang
  et~al\mbox{.}}{2017}]%
        {ouyang2017fast}
\bibfield{author}{\bibinfo{person}{Peng Ouyang}, \bibinfo{person}{Shouyi Yin},
  {and} \bibinfo{person}{Shaojun Wei}.} \bibinfo{year}{2017}\natexlab{}.
\newblock \showarticletitle{A fast and power efficient architecture to
  parallelize LSTM based RNN for cognitive intelligence applications}. In
  \bibinfo{booktitle}{\emph{Proceedings of the 54th Annual Design Automation
  Conference 2017}}. ACM, \bibinfo{pages}{63}.
\newblock


\bibitem[\protect\citeauthoryear{Quinn}{Quinn}{2003}]%
        {quinn2003parallel}
\bibfield{author}{\bibinfo{person}{Michael~J Quinn}.}
  \bibinfo{year}{2003}\natexlab{}.
\newblock \showarticletitle{Parallel Programming}.
\newblock \bibinfo{journal}{\emph{TMH CSE}}  \bibinfo{volume}{526}
  (\bibinfo{year}{2003}).
\newblock


\bibitem[\protect\citeauthoryear{Sak, Senior, and Beaufays}{Sak
  et~al\mbox{.}}{2014}]%
        {sak2014long}
\bibfield{author}{\bibinfo{person}{Ha{\c{s}}im Sak}, \bibinfo{person}{Andrew
  Senior}, {and} \bibinfo{person}{Fran{\c{c}}oise Beaufays}.}
  \bibinfo{year}{2014}\natexlab{}.
\newblock \showarticletitle{Long short-term memory recurrent neural network
  architectures for large scale acoustic modeling}. In
  \bibinfo{booktitle}{\emph{Fifteenth annual conference of the international
  speech communication association}}.
\newblock


\bibitem[\protect\citeauthoryear{Shin and Sung}{Shin and Sung}{2016}]%
        {shin2016dynamic}
\bibfield{author}{\bibinfo{person}{Sungho Shin} {and} \bibinfo{person}{Wonyong
  Sung}.} \bibinfo{year}{2016}\natexlab{}.
\newblock \showarticletitle{Dynamic hand gesture recognition for wearable
  devices with low complexity recurrent neural networks}. In
  \bibinfo{booktitle}{\emph{Circuits and Systems (ISCAS), 2016 IEEE
  International Symposium on}}. IEEE, \bibinfo{pages}{2274--2277}.
\newblock


\bibitem[\protect\citeauthoryear{Sundermeyer, Schl{\"u}ter, and
  Ney}{Sundermeyer et~al\mbox{.}}{2012}]%
        {sundermeyer2012lstm}
\bibfield{author}{\bibinfo{person}{Martin Sundermeyer}, \bibinfo{person}{Ralf
  Schl{\"u}ter}, {and} \bibinfo{person}{Hermann Ney}.}
  \bibinfo{year}{2012}\natexlab{}.
\newblock \showarticletitle{LSTM neural networks for language modeling}. In
  \bibinfo{booktitle}{\emph{Thirteenth Annual Conference of the International
  Speech Communication Association}}.
\newblock


\bibitem[\protect\citeauthoryear{Tang, Qin, and Liu}{Tang
  et~al\mbox{.}}{2015}]%
        {tang2015document}
\bibfield{author}{\bibinfo{person}{Duyu Tang}, \bibinfo{person}{Bing Qin},
  {and} \bibinfo{person}{Ting Liu}.} \bibinfo{year}{2015}\natexlab{}.
\newblock \showarticletitle{Document modeling with gated recurrent neural
  network for sentiment classification}. In
  \bibinfo{booktitle}{\emph{Proceedings of the 2015 conference on empirical
  methods in natural language processing}}. \bibinfo{pages}{1422--1432}.
\newblock


\bibitem[\protect\citeauthoryear{Turchenko, Bosilca, Bouteiller, and
  Dongarra}{Turchenko et~al\mbox{.}}{2013}]%
        {turchenko2013efficient}
\bibfield{author}{\bibinfo{person}{Volodymyr Turchenko},
  \bibinfo{person}{George Bosilca}, \bibinfo{person}{Aurelien Bouteiller},
  {and} \bibinfo{person}{Jack Dongarra}.} \bibinfo{year}{2013}\natexlab{}.
\newblock \showarticletitle{Efficient parallelization of batch pattern training
  algorithm on many-core and cluster architectures}. In
  \bibinfo{booktitle}{\emph{Intelligent Data Acquisition and Advanced Computing
  Systems (IDAACS), 2013 IEEE 7th International Conference on}},
  Vol.~\bibinfo{volume}{2}. IEEE, \bibinfo{pages}{692--698}.
\newblock


\bibitem[\protect\citeauthoryear{Weste and Eshraghian}{Weste and
  Eshraghian}{1994}]%
        {weste1994principles}
\bibfield{author}{\bibinfo{person}{Neil~HE Weste} {and}
  \bibinfo{person}{Kainran Eshraghian}.} \bibinfo{year}{1994}\natexlab{}.
\newblock \showarticletitle{Principles of CMOS VLSI design: A systems
  perspective second edition}.
\newblock \bibinfo{journal}{\emph{Addision-Wesley Publishing, California,
  l994}} (\bibinfo{year}{1994}).
\newblock


\bibitem[\protect\citeauthoryear{Xianyi, Qian, and Chothia}{Xianyi
  et~al\mbox{.}}{2012}]%
        {xianyi2012openblas}
\bibfield{author}{\bibinfo{person}{Zhang Xianyi}, \bibinfo{person}{Wang Qian},
  {and} \bibinfo{person}{Zaheer Chothia}.} \bibinfo{year}{2012}\natexlab{}.
\newblock \showarticletitle{OpenBLAS}.
\newblock \bibinfo{journal}{\emph{URL: http://xianyi. github. io/OpenBLAS}}
  (\bibinfo{year}{2012}), \bibinfo{pages}{88}.
\newblock


\end{thebibliography}

\end{document}